\documentstyle[prb,aps,epsf,twocolumn]{revtex}
\begin{document}

\hfuzz=100pt
\hbadness=10000

\twocolumn[\hsize\textwidth\columnwidth\hsize\csname
@twocolumnfalse\endcsname

\title{\bf Low temperature shape relaxation of 2-d islands by edge
  diffusion}

\author{Nicolas Combe (a), Hern\'an Larralde
  (b)}
  \address{(a) D\'epartement de Physique
    des
    Mat\'eriaux, UMR CNRS 5586, Universit\'e Claude Bernard Lyon-1,
    69622
    Villeurbanne Cedex, FRANCE\\ (b) Centro de Ciencias
    F\'{\i}sicas,
    U.N.A.M., Apdo. Postal 48-3, C.P. 62251, Cuernavaca, Morelos,
    MEXICO}

    \maketitle

\begin{abstract} 

We present a precise microscopic description of the limiting step for
low temperature shape relaxation of two dimensional islands in which
activated diffusion of particles along the boundary is the only
mechanism of transport allowed. In particular, we are able to explain
why the system is driven irreversibly towards equilibrium. Based on
this description, we present a scheme for calculating the duration of
the limiting step at each stage of the relaxation process. Finally, we
calculate numerically the total relaxation time as predicted by our
results and compare it with simulations of the relaxation process.
\end{abstract}

\vskip2pc]
\narrowtext

\section{Introduction}

The understanding, description and control of structures at the
nanometer scales is a subject of interest from the fundamental and
applied points of view \cite{generale,revue}. From the fundamental
point of view, there is a large literature \cite{noziere,pimpinelli}
concerning the growth of crystals and their shape. Yet, while the
description of the equilibrium shape is rather clear, the dynamic
description of crystal growth is still not well understood. In
particular, we lack a complete understanding of the time scales
involved in the relaxation process, and the mechanisms which
irreversibly conduce the island to its equilibrium shape.

In this work, we study the shape relaxation of two dimensional islands
by boundary diffusion at low temperatures. The typical size of the
islands we will be concerned with consists of a few thousand atoms or
molecules, corresponding to islands of a few nanometers. The 
model we consider is the same as the one studied in \cite{eur_physB},
where two mechanisms of relaxation, depending on temperature, were
pointed out: At high temperatures, the classical theory developed by
Herring, Mullins and Nichols \cite{nichols} appears to describe
adequately the relaxation process. In particular, it predicts that the
relaxation time scales as the number of atoms to the power
$2$. However, at low temperatures, the islands spend long
times in fully faceted configurations, suggesting that the limiting
step of the relaxation in this situation is the nucleation of a new
row on a facet. This assumption leads to the correct scaling
behavior of the relaxation time on the size of the island, as well as
the correct temperature dependence. Yet, it is unclear what drives 
the island towards equilibrium in this scenario.

In this paper we propose a detailed description of this low
temperature relaxation mechanism, and identify the event that drives
the island towards its equilibrium shape. Based on our description, we
construct a Markov process from which we can estimate the duration of
each stage of the relaxation process. Finally, we use our result to
determine the relaxation time of the islands and compare with
simulation results.

The specific model under consideration consists of 2D islands having a
perfect triangular crystalline structure. A very simple energy
landscape for activated atomic motion was chosen: the aim being to
point out the basic of mechanisms of relaxation, and not to fit the
specific behavior of a particular material.  The potential energy
$E_p$ of an atom is assumed to be proportional to the number $i$ of
neighbors, and the {\it kinetic barrier} $E_{act}$ for diffusion
is also proportional to the number of {\it initial} neighbors before
the jump, regardless of the {\it final} number of neighbors: $
E_{act}=- E_p = i*E $ where $E$ sets the energy scale ($E=0.1$ eV
throughout the paper). Therefore, the probability $p_i$ per unit time
that an atom with $i$ neighbors moves is $p_i = \nu_0
\exp[-i*E/k_bT]$, where $\nu_0= 10^{13} s^{-1}$ is the Debye
frequency, $k_B$ is the Boltzmann constant and $T$ the absolute
temperature. Hence, the average time in which a particle with $i$
neighbors would move is given by :
\begin{equation} 
\tau_i=  \nu_0^{-1} \exp[i*E/k_bT]
\label{taui} 
\end{equation}
The complete description of the model and of the simulation algorithm
can be found in \cite{eur_physB}, where it was studied using Standard
Kinetic Monte Carlo simulations. This simple kinetic model has only
{\it one} parameter, the ratio $E/k_B T$. The temperature was varied
from $83 K$ to $500 K$, and the number of atoms in the islands from
$90$ up to $20000$. The initial configurations of the islands were
elongated (same initial aspect ratio of about 10), and the simulations
were stopped when the islands were close to equilibrium, with an
aspect ratio of 1.2. The time required for this to happen was defined
as the relaxation time corresponding to that island size and
temperature.
Concerning the dependence of the relaxation time on the size of the
island, two different behaviors depending on temperature were
distinguished\cite{eur_physB}. At high temperature, the relaxation
time scaled as the number of atoms to the power $2$, but this exponent
decreased when temperature was decreased.  A careful analysis showed
that the exponent tends towards $1$ at low temperature.  The
dependence of the relaxation time as a function of temperature also
changes, the activation energy was 0.3 eV at high temperature and 0.4
eV at low temperature. In this context, it is important to define what
we call a low temperature: following \cite{eur_physB}, we denote by
$L_c$ the average distance between kinks on a infinite facet: we
define the low temperature regime as that in which $L_c \gg L$ where L
is the typical size of our island, large facets are then visible on
the island. It was shown that $L_c=\frac{a}{2} exp(\frac{E}{2k_bT})$
where $a$ is the lattice spacing.\\
The behavior of the relaxation time as a function of the temperature
and $N$, the number of particles of the island, can be summed up with
two equations corresponding to the high and low temperature regimes:
\begin{eqnarray} 
t^{HT}_{relaxation}& \propto &  \exp[3E/k_bT] N^2 \;  \mbox{for} \; N \gg L_c^2 \label{teqHT} \\
t^{LT}_{relaxation}& \propto & \exp[4E/k_bT] N \; \mbox{for} \; N \ll L_c^2 \label{teqLT}
\end{eqnarray} 
Replacing the temperature dependent factors by a function of $N_c$ the
crossover island size (where $N_c=L_c^2 \propto exp(E/k_bT)$), these
two laws can be expressed as a unique scaling function depending on
the rescaled number of particles $N/N_c$:
\[ t_{relaxation } \propto \left\{ 
\begin{array}{ll}
   N_c^{5} \left(\frac{N}{N_c}\right)^2 \;
& \mbox{for} \;\frac{N}{N_c} \gg 1  \\
  N_c^{5} \frac{N}{N_c} \;
 & \mbox{for} \;\frac{N}{N_c} \ll 1 
\end{array} \right. \]
So that the relaxation time \cite{note} is a simple monotonous function of
$N/N_c$, and the temperature dependence is contained in $N_c$. \\
We will now focus on the precise microscopic description of the limiting
step for relaxation in the low temperature regime.

\section{Description of the limiting process at low temperature}

\subsection{Qualitative description}

\begin{figure}
\centerline{\epsfxsize=5cm
\epsfbox{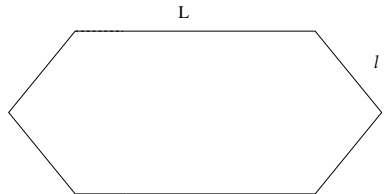}
}
\caption{Configuration of the island we will consider in the qualitative description as well as in the quantitative one.}
\label{island} 
\end{figure}

During relaxation at low temperature, islands are mostly in fully
faceted configurations. Let us, for instance, consider an island in a
simple configuration given by fig.~\ref{island}. When $L$ is larger
than $l$, the island is not in its equilibrium shape (which should be
more or less a regular hexagon). To reach the equilibrium shape,
matter has to flow from the ``tips'' of the island (facets of length
$l$ in this case) to the large facets $L$. In this low temperature
regime there are very few mobile atoms at any given time, therefore
this mass transfer must be done step by step: the initial step being
the nucleation of a ``germ'' of two bound atoms on a facet of length
$L$ and then, the growth of this germ up to a size $L-1$ due to the
arrival of particles emitted from the kinks and corners of the
boundary of the island. Thus the germ grows, and eventually completes
a new row on the facet.

This simple picture still leaves a basic question unanswered: the
relatively faster formation of a new row on a small facet would lead
the island further away from its equilibrium shape, and yet, we
observe that this never happens. Indeed, sometimes a germ appears on a
small facet but it eventually disappears afterwards, whereas the
appearance of a germ on a large facet frequently leads to the
formation of a new row, taking the island closer to its equilibrium
shape.

These observations are at the root of irreversible nature of the
relaxation, germs only grow and become stable on the large facet, so
the island can only evolve to a shape closer to equilibrium. Yet,
there is clearly no local drive for growth on large facets nor any
mechanism inhibiting growth on small ones. In order to explain how
this irreversibility comes about, we propose the following detailed
description of the mechanism of nucleation and of growth of a germ.

\begin{figure}
\centerline{\epsfxsize=5cm
\epsfbox{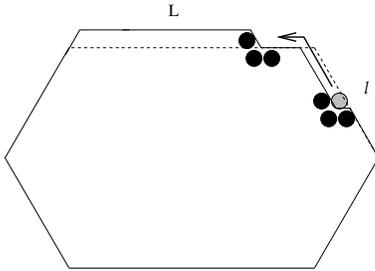}
}
\caption{The total energy of the island does not change when one particle leaves a kink on the small facet to go in the germ on the large facet.}
\label{island2} 
\end{figure}

First, to create a germ, 2 atoms emitted from the corners of the
island have to encounter on a facet. The activation energy required
for this event is obviously independent of whether it occurs on a
large facet or on a small facet. Once there is a germ of 2 atoms on a
facet, the total energy of the island {\it does not} change when a
particle is transfered from a kink to the germ (3 bonds are broken,
and 3 are created) see Fig.~\ref{island2}. Clearly the same is true if
a particle from the germ is transfered to its site of emission or any
other kink. Thus, germs can grow or decrease randomly without energy
variations driving the process. An exception to this occurs if the
particle that reaches the germ is the last one of a row on a facet; in
that case, the energy of the system decreases by $1$ binding energy
E. The island is then in a configuration from which it is extremely
improbable to return to the previous configuration. For this to occur,
a new germ would have to nucleate (and grow) on the original facet.
This event is almost impossible in the presence of the kinks of the
first growing germ, which act as traps for mobile atoms.  Thus, when a
germ nucleates on a facet, it can grow or decrease without changing
the energy of the island except if a complete row on a facet
disappears, in which case it ``stabilizes''.
The scenario above explains why no new rows appear on small facets:
when a germ grows on a small facet, since atoms come either from a
small or a large facet, no complete row of a facet can disappear
during the germ's growth, and thus, the island never decreases its
energy.  On the other hand when the germ grows on a large facet, the
germ might grow or decrease, but if the size of the germ reaches the
size of the small facet, the energy of the system will decrease and
the system has almost no chance to go back its previous shape. We
believe that this is the microscopic origin of irreversibility in the
relaxation of this system.  It should be stressed that this scenario
for the growth and stabilization of germs is different from usual
nucleation theory, where the germ has to overcome a free energy
barrier \cite{nucleation} to become stable.

This microscopic description also shows that the limiting step for
this ``row by row'' relaxation mechanism is actually the formation on a
large facet, of a germ of the size of the small facet. This fact
allows us to estimate the duration of the limiting step at each stage
of the relaxation process.

\subsection{A Quantitative description}
\label{quanti_description}

Based on our description of the process, we propose a scheme for
calculating the time required to form a stable germ, i.e. a germ of
size $l$, on a facet of size L. As mentioned above, the appearance of
this stable germ is the limiting step for the formation of a new row
on that facet.

The idea is to describe the growth of the germ as a succession of
different island states, and calculate the probability and the time to
go from one state to another in terms of the actual diffusive
processes occurring on the island surface. These states form a Markov
chain, the future evolution of the system being essentially determined
by the state of the system, independently of its previous behavior.

As a further simplification, we consider a simple fully faceted
island in a elongated hexagonal shape whose facets are of length $L$
and $l$, see Fig.~\ref{island}; moreover, we normalize every length by
the lattice spacing $a$.

\begin{figure}
\centerline{\epsfxsize=5cm
\epsfbox{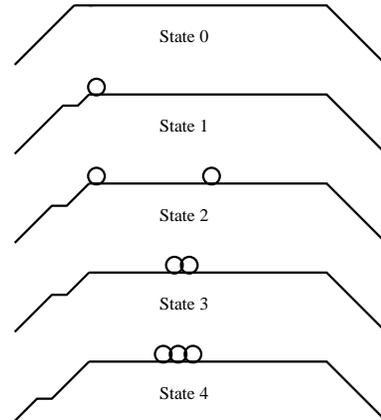}
}
\caption{Outline of the states considered in the Markov chain.}
\label{diff_state} 
\end{figure}

The different states we consider are (See Fig.~\ref{diff_state} also)
:
\\
- state $0$ : there is no particle on the facets.\\
- state $1$ : one particle is on one end of a facet $L$. \\
- state $2$ : 2 particles are on the facet $L$: one of them is on one
end of the facet,
and the other one has diffused from an end.\\
- state $3$ : 2 particles are on the facet $L$ but they are bonded together.\\
- state $4$ : 3 bonded particles are on the facet $L$.\\
\ldots \\
- state $n$ : $n-1$ bonded particles are on the facet $L$.\\

The goal of this calculation is to estimate the time to go from state
$0$ to state $l+1$. We treat the problem as a discrete time Markov chain. The
unit time being $\tau_2$, the typical time for a particle with 2
neighbors to move. This time is in fact the smallest relevant time of
the system so that the operation of discretization does not affect the
results.  In the following, the time $\tau_i$ is the average time for
a particle with $i$ neighbors to move: $\tau_i=\nu_0^{-1}
e^{\frac{i*E}{kT}}$.  For clarity, we will use the term {\it time} for
the discrete time of the Markov chain, and the term {\it real time}
for the time of physical process. The obvious relation between the two
time scales is : $time= \frac{real\ time}{\tau_2}$.\\

We define the parameter $\rho = \frac{\tau_2}{\tau_3}$. In the limit
of small temperature, $\rho$ is a very small quantity. Moreover, one
can easily check that $\rho= exp(-\frac{E}{kT})= \frac{2}{L_c^2}$
where $L_c$ is the average distance between kinks defined in the first
section. So that the condition $L_c \gg L$ (low temperature regime)
could be written as: $\sqrt{\rho} L \ll 1$ or $\rho L \ll 1/L < 1$.
\\

Denote by $\alpha_i$ the probability for the system in state $i$ to
stay in the state $i$ the following step, $p_i$ the transition
probability for system in state $i$ to go to state $i+1$, and $q_i$
the transition probability for the system to go into state $i-1$. \\
We have now to evaluate the different quantities $\alpha_i,~p_i$ and
$q_i$ in terms of the diffusive processes that take place on the
island's boundary.

We first evaluate the quantities $p_1,q_1$ and $\alpha_1$.  Let us
assume that the states $0$ and $2$ are absorbent; the average time
$n_1$ needed to leave state $1$, corresponds to the average real time
a particle stays on a facet starting on one of its edges, which act as
traps. Since the particle performs a random walk, this can be readily
calculated to be $L \tau_2$. So we should have :
\begin{equation}
n_1=\frac{1}{1-\alpha_1}=L
\end{equation} 

Moreover, the probability to go to state $2$ is the probability that a
new particle leaves a kink and reaches the facet while the first
particle is still on the facet, we calculate this probability in
Appendix~\ref{appen_1} where we find:
\begin{equation} 
P = 1 - \left[\frac {2 \sinh(2 \sqrt{\rho} (L-1)) }{\sinh(2\sqrt{\rho}L) } -
\frac {\sinh(2\sqrt{\rho} (L/2-1)) }{\sinh(2\sqrt{\rho}L/2) } \right]
\label{P_exact_text}.
\end{equation}
We expand expression Eq.~\ref{P_exact_text} for small $\rho$ keeping the
first term :
\begin{eqnarray} 
P = 2 \rho (L-1) + o(\rho)\label{appen_1_final}
\end{eqnarray} 
We could also calculate the probability $b_1$, that the 
system in state $1$ eventually reaches state $2$ as:
\begin{equation}
b_1=\alpha_1 b_1 + p_1.
\end{equation}
Thus $b_1 = \frac{p_1}{1-\alpha_1}$, and using
$p_1+\alpha_1+q_1 = 1 $ we find:
\begin{eqnarray}
p_1&=& \frac{P}{L} \simeq  2 \rho \frac{L-1}{L} + o(\rho)  \label{p1}\\
\alpha_1 &=& 1 - 1/L \label{a1}\\
q_1 &=& \frac{1}{L} - \frac{P}{L} \simeq \frac{1}{L} - 2 \rho \frac{L-1}{L} + o(\rho) \label{q1} 
\end{eqnarray}   
We can do the same with state $2$, knowing that the probability for
two particles to stick (state $3$) is $\lambda/L$, where $ \lambda =
\left(\frac{\cosh \pi +1 } { \sinh \pi} \right)\pi$ (the calculation
of this probability is carried out in appendix~\ref{appen_11}). The
time to leave state $2$ is $\kappa L$, where $\kappa$ is a numerical
constant given by $ \kappa = \frac{4}{\pi^2} \sum_{k=1}^{\infty}
\frac{1}{1+4k^2} \left[ 2 - \frac{1}{(2k+1)} \right]$ (see
appendix~\ref{appen_12}). Thus, we obtain:
\begin{eqnarray} 
 p_2 &=&  \frac{\lambda}{\kappa L^2} \label{p2}\\
\alpha_2 &=& 1 - \frac{1}{\kappa L} \label{a2}\\
q_2 &=& \frac{1}{\kappa L} - \frac{\lambda}{\kappa L^2} \label{q2}
\end{eqnarray} 

In order to obtain a chain which can be treated analytically, we
assume that the probability to go from state 3 to state 4, is the same
as the probability to go from state 4 to state 5 and, in general, that
the probabilities to go from state $i$ to state $i+1$ are $p_i =
p_{i+1}=p$ for $i \ge 3$. Similarly, we assume that
$\alpha_i=\alpha_{i+1}= \alpha$ for $i \ge 3$ and $q_i=q_{i+1} = q$
for $i > 3$.

To calculate the probabilities $p,q,\alpha$, in appendix~\ref{appen_2}
we have calculated the average real time $t_{p,q,\alpha}$ to go from
state $i$ to state $i+1$ assuming the average distance between kinks
and the germ is $L/2$:
$t_{p,q,\alpha}=L/4 *\tau_3+ L(L-2)/4*\tau_2$. Moreover, since $p=q$ ,
we can calculate $p,q$ and $\alpha$ :
\begin{eqnarray}
p &=& 2 \rho / L - \frac{2(L-2)}{L}\rho^2 + o(\rho^2)\label{pk}\\
\alpha &=& 1 - \frac{4 \rho}{L}+ \frac{4(L-2)}{L}\rho^2 + o(\rho^2) \label{ak} \\
q &=& 2 \rho / L - \frac{2(L-2)}{L}\rho^2 + o(\rho^2)\label{qk}
\end{eqnarray}

When 2 particles are bonded on the facet (state 3), the probability to
go to state $2$ should practically be equal to $q$, we will assume
this to be the case.

So far, we have omitted the possibility that the germ can also
nucleate on a small facet $l$. To take this into account, we
have to consider new states : \\
- state $-1$: one particle is on the facet $l$ on one of its edges. \\
- state $-2$: 2 particles are on the facet $l$: one of them is on
               the edge of the facet, and the other has diffused from an 
               edge\\
- state $-3$: 2 particles are on the facet $l$ but they are bonded.\\
- state $-4$: 3 bonded particles are on the facet $l$.\\
\ldots \\                   
- state $-l$: $l-1$ bonded particles are on the facet $l$.\\

As discussed above, if the system arrives to state $-l$, a row on a
small row is completed, which is not an absorbent state, and since
this row cannot grow any further the system can only go
back to state $-l+1$ or stay in state $-l$. 

The different probabilities of transition from one state to another in
this branch of the chain are the same as the ones calculated before,
replacing $L$ by $l$. Thus we have $q_{-i}=p_{i}(L \Rightarrow l)$,
$p_{-i}=q_{i}(L \Rightarrow l)$ and $\alpha_{-i}=\alpha_{i}(L
\Rightarrow l)$ for $i \geq 1$ except for the state $-l$, where we
always have $\alpha_{-i}=\alpha(L \Rightarrow l)$, but, $p_{-l}=2*q(L
\Rightarrow l)$. In the following, we will use the notation :
$p_i^*=p_{i}(L \Rightarrow l)$, the values of $p_i$ where we have
replace $L$ by $l$.

To complete the calculation, we now have to determine $p_0,q_0$ and
$\alpha_0$.  The average real time the system stays in state $0$,
assuming states $1$ and $-1$ are absorbent, is almost $\tau_3/2$ if we
take into account that there are {\it two} kinks, one at each end of
the facet. The probability that the germ nucleates on the facet
$L$ is simply $\frac{L}{L+2l}$. From this we deduce:
\begin{eqnarray} 
p_0&=&2\frac{L}{L+2l} \rho \label{p0}\\ 
\alpha_0&=&1-2\rho \label{a0} \\
q_0&=&2\frac{2l}{L+2l} \rho \label{q0}
\end{eqnarray}

\begin{figure}
\centerline{\epsfxsize=8cm
\epsfbox{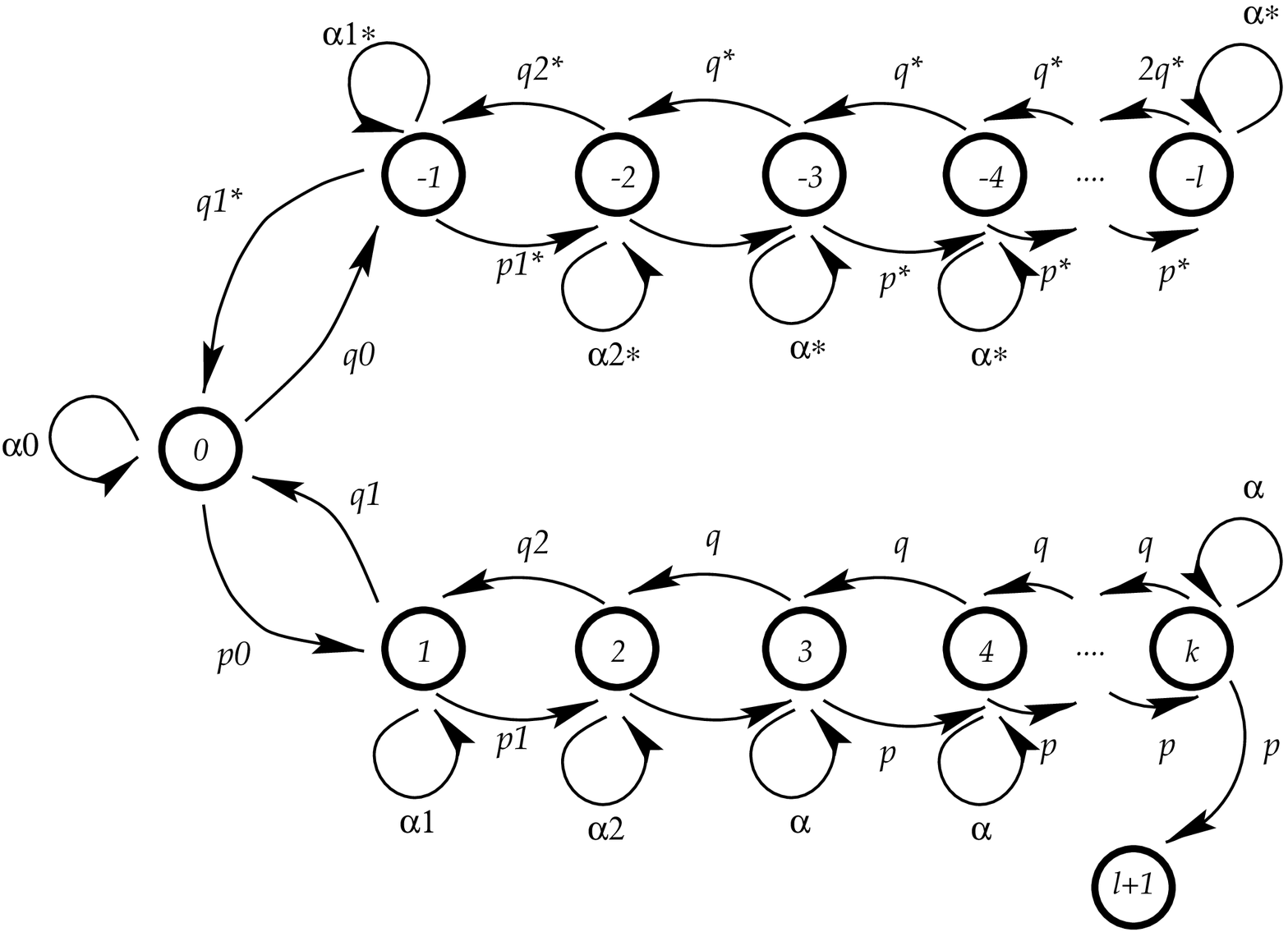}
}
\caption{Diagram of the entire Markov chain with the two branches :
 the upper one for the germs growing on a small facet and the other
 one for the germs growing on the large one, state $l+1$ is the only absorbing
 state.}
\label{chaine2} 
\end{figure}

The diagram of the entire Markov chain is then given by
Fig.~\ref{chaine2}. And we will calculate the time to go from state
$0$ to state $l+1$.\\

The state $l+1$ is absorbent (as discussed above, when the size of the
germ reaches {\it l } on a large facet, the system cannot go back to
the initial state). Let us call $n_i$ the average time to go from
state $i$ to state $l+1$. We can write :
\begin{eqnarray}
n_{-l}&=& 1+ 2q^* n_{-l-1} + \alpha^* n_{-l} \label{eq-l} \\
n_{-k} &=& 1 + q^* n_{-k-1} + \alpha^* n_{-k} + p^* n_{-k+1} \label{eq-k}\\
with & & 3 \le k \le l \\
&...& \nonumber \\ 
n_{-2} &=& 1 + q_2^* n_{-1} + \alpha_2^* n_{-2} + p_2^* n_{-3} \label{eq-2}\\
n_{-1} &=& 1 + q_1^* n_0 + \alpha_1^* n_{-1} + p_1^* n_{-2} \label{eq-1} \\
n_0 &=& 1 + q_0 n_{-1} + \alpha_0 n_0 + p_0 n_1 \label{eq0} \\
n_1 &=& 1 + q_1 n_0 + \alpha_1 n_1 + p_1 n_2 \label{eq1} \\
n_2 &=& 1 + q_2 n_1 + \alpha_2 n_2 + p_2 n_3 \label{eq2}\\
&...& \nonumber  \\
n_k &=& 1 + q n_{k-1} + \alpha n_k + p n_{k+1} \label{eqk}\\
 with & & 3 \le k \le l-1 
\end{eqnarray}
The boundary condition for this process is $n_{l+1}=0$. The
calculation of $n_0$ is quite straight forward but tedious, it is
carried out in the appendix~\ref{appen_3}. We find that in the limit of small
temperatures, the typical real time needed to nucleate a germ of size
$l$ on a facet of size $L$ is given by:
\begin{equation} 
\tau(L,l) \approx \frac{\tau_3^2}{\tau_2} \frac{(L+2l)}{4L(L-1)} 
\left[ (\frac{L}{\lambda}-1)(l-1)+1 \right] 
\end{equation} 
where we have kept only the most relevant term at low temperatures.

\section{Estimation of the Relaxation time} 

\vspace{.4cm}
{\it Scaling laws}
\vspace{.4cm}

In this section we calculate the typical time required for an island
to relax from an initial out of equilibrium shape. We assume that at
all times, the instantaneous shape of the island can be characterized
by the lengths $L$ and $l$ of its long and short facets respectively.
Then, following the discussion in the previous sections, a new row of
particles will appear on a long facet after a time $\tau(L,l)$. Thus,
calling $v(L)$ the normal speed of the large facet and taking the
particle size as our unit distance, we have:

\begin{equation} 
v(L) \approx \frac{1}{\tau(L,l)} \approx \frac{4 \tau_2 L
(L-1)}{\tau_3^2 (L+2l)\left[ (\frac{L}{\lambda}-1)(l-1)+1 \right]}
\label{v(l)}
\end{equation}

The scaling properties of the relaxation time can be deduced by
noticing that the length scales involved scale as $N^{1/2}$, where N
is the number of atoms of the island. Thus we renormalize the lengths
by $x \rightarrow x'= N^{-1/2} x$. Then, to scale out the size
dependence as $N$ grows, one must rescale time by $ t \rightarrow t'=t
N^{-1}$.  This is the result obtained in \cite{eur_physB}: at low
temperature, the relaxation time is proportional to the number of the
atoms of the island. But, as we expect our results for the time
required to complete a row at each stage, as given in
Eq.~\ref{n0_exact}, to be relatively accurate, we can go beyond the
scaling properties and use it to calculate numerically the time
required for the complete relaxation process, including the
corrections arising from the lower order terms.

In what follows, we establish the differential equations which permit
the calculation of the full relaxation time of an island.  As
mentioned above, we still consider the simple island of
Fig.~\ref{island}, where $v(L)$ is the normal speed of the facet $L$,
and $v(l)$ that of the facet $l$.  We now consider $L$ and $l$ as
continuous variables, which considerably simplifies the calculation.

\begin{figure} 
\centerline{ \epsfxsize=8cm
\epsfbox{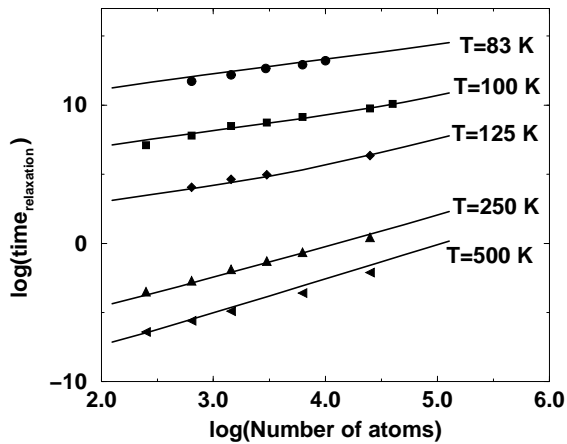} }
\caption{Size dependence of the relaxation time for different temperatures in log-log bases. Filled points have been obtained by Monte Carlo Simulations whereas the solid lines have been obtained by integration
of the system~\ref{eq_diff1},~\ref{eq_diff2}. The agreement of the two analysis is almost perfect.}
\label{integ} 
\end{figure}

Conservation of the matter imposes the relation:
\begin{equation} 
Lv(L)+2lv(l)=0
\end{equation} 
Moreover, we can find geometric relations between $L,l,v(L),v(l)$ : 
\begin{eqnarray} 
v(L)&=&\sqrt{3}/2  \frac{dl}{dt} \\
v(l)&=&\sqrt{3}/4(\frac{dl}{dt} + \frac{dL}{dt} )
\end{eqnarray}
So finally we find: 
\begin{eqnarray}
\frac{dl}{dt}&=&\frac{2}{\sqrt{3}} \frac{1}{\tau(L,l)} \label{eq_diff1}\\
\frac{dL}{dt}& =& - \frac{2}{\sqrt{3}} \left( \frac{L}{l}+1 \right) \frac{1}{\tau(L,l)} 
\label{eq_diff2}
\end{eqnarray} 
To integrate these equations numerically: we use Eq.~\ref{n0_exact},
and for $p1,q1$ and $\alpha_1$ the exact estimation using
Eq.~\ref{P_exact}, as well as the explicit values of the quantities
$q_i,\alpha_i$ and $p_i$ we have found: Eqs. [\ref{p2}-\ref{q0}]. We
start the integration from an island of aspect ratio of $R=10$, and
stop it when the aspect ratio is $R=1.2$. Aspect ratios are explicitly
calculated as:
\begin{eqnarray}
R &=& \frac{r_x}{r_y} \\
 r_x^2 &=&\frac{1}{S} \int \!\!\!\! \int_{Island \ Surface} (x-x_G)^2 dx dy \\
 r_y^2 &=&\frac{1}{S}a
\int \!\!\!\! \int_{Island \ Surface} (y-y_G)^2 dx dy \\
 S &=& \int \!\!\!\! \int_{Island \ Surface} dx dy
\end{eqnarray} 
Where $x_G$ and $y_G$ give the position of the center of gravity of
the island.  We report in Fig.~\ref{integ} the relaxation time as a
function of $N$, the number of particles of the island in a log-log plot.
 
We find a good quantitative agreement between the simulations and our
predictions except at the highest temperatures (see below). 

\section{Summary and Discussion} 

We have considered the shape evolution of two dimensional islands as a
result of the nucleation of a germ on a facet which then grows or
decreases due to single particle processes. Then we recognized that
the disappearance of a complete row is responsible for the
stabilization of the germ, and that this is only feasible for germs
growing on the large facets. This gives rise to an overall flux of
particles from the small facets to the large facets which leads the
shape of the island irreversibly towards equilibrium. Based on this
description, we have recast the formation of stable germs, which is
the limiting step for relaxation, into a Markov chain in which the
transition probabilities are calculated in terms of the underlying
diffusive processes taking place on the island's boundary. Solving
this chain yields an estimate of the time of formation of a new row at
each stage of the relaxation. Integrating our results we can obtain
the relaxation time for the evolution of an island from an aspect
ratio of 10 to an aspect ratio of 1.2, as a function of temperature
and island size. Our results have a rather good quantitative agreement
with those obtained from direct simulations of the system.  At higher
temperatures, multiple nucleation processes, the presence of many
mobile particles on the island's boundary and the failure of our
hypothesis $L_c \gg L$ (Sect.~\ref{quanti_description}) invalidate our
picture and the relaxation becomes driven by the coarsed grained
curvature of the boundary, which leads to Mullins classical theory.

The description of the simple model considered in this work is
certainly not exact and there are other effects that could be taken
into account. Perhaps the most important effect we have overstepped at
low temperatures, is related to our assumption that after the germ
stabilizes a single new row is formed on the large facet. The
differential equations for the evolution of the island were derived
from this assumption. It is clear that this is not correct: our
estimation of the time required to stabilize a germ starts from a
fully faceted configuration, and once a complete row on a small facet
disappears the germ becomes stable and a full row on the large facet
can be formed. Once this row is finished, it is very unlikely that the
island will be in a fully faceted configuration again, leaving at
least an extra kink on the boundary. This gives rise to extra sources
and traps for mobile particles, which might have affected the relaxation
rate. Another issue is our characterization of the faceted island with
only two facets sizes: a more detailed characterization may be
relevant especially in the early stages of relaxation.
It is clear that more accurate models for specific systems can also be
constructed. These could take into account the dependence of the
edge-diffusion coefficients on the orientation of the facet, as well
as the dependence of emission rates on the local geometry. Such
dependences have been studied, for example, by \cite{ferrando}. In
terms of the elements of description we use, inclusion of these
effects would be achieved by changing the values of $\tau_2$
(diffusion time) and $\tau_3$ (emission time from kinks and corners)
depending on the orientation of the facets involved in each
event. Thus, the nucleation time would depend on the facet upon which
it happens. Such a dependence of the nucleation time may drive the
island toward a non regular hexagon equilibrium shape
\cite{michely}, and would reproduce the phenomenology of a larger variety of
materials. These changes will affect the temperature dependent
prefactors in our results, as these depend on the temperature through
$\tau_2$ and $\tau_3$. However, the size dependence of the nucleation
time and of the relaxation time, which is where the departure from
Mullin's theory is evidenced, would stay the same.\\
From a more general point of view, only the diffusion of particles along
the perimeter of islands has been taken into account in this work. In
real systems, other mechanisms can contribute to the transport of the
matter which leads to relaxation: volume diffusion and transport
through the two dimensional gas of particle surrounding the island.
Volume diffusion is usually a much slower process than the other two,
and can usually be neglected safely. On the other hand, it is well
known that edge diffusion is more efficient for short trips whereas
transport through the 2D gas is faster on long distances.  Following
Pimpinelli and Villain in \cite{pimpinelli}(p.132), a characteristic
length $r_1$ beyond which edge diffusion is less efficient than
transport through the gas can be evaluated : $r_1 \approx \sqrt{D_s
\tau_v}$ where $D_s$ is the edge diffusion coefficient and $1/\tau_v$
the probability per unit time a given particle leaves the island. So
that our assumptions should be valid for islands with a number N of
particles such that : $N \ll D_s \tau_v$. Since the activation energy
for edge diffusion is smaller than the activation energy for
evaporation, $r_1$ is a decreasing quantity with temperature, and we
expect $r_1$ to be very large at low temperature. Thus, this mechanism
is essentially irrelevant in the description of the evolution of
nanometer structures at low temperatures. Moreover, recent
experimental results \cite{stoldt} have shown that supported Ag two
dimensional islands relax via atomic diffusion on the island
perimeter, without significant contribution from exchange with the two
dimensional gas.\\
Finally, our results are to be compared with a recent theoretical study
\cite{PRL} concerning the relaxation of three dimensional
crystallites.  This study also points out two relaxation regimes as a
function of temperature.  At high temperature the relaxation scales in
accordance with the results derived from Mullins' theory, whereas at
low temperature the relaxation time becomes an exponentional function
of the size of the crystallites. So that the effects of lowering the
temperature are qualitatively different for two dimensional and three
dimensional crystallites : in two dimensions, lowering the temperature
decreases the strength of the dependence of the relaxation time as a
function of the size of the crystallites (as it crosses over from a
$N^2$ dependence to a $N$ dependence), whereas it increases this the
strength in three dimensions. In both cases, the limiting step is the
nucleation of a germ on a facet : a unidimensional germ in two
dimensions, and a two dimensional germ in three dimensions. The
difference stems from the fact that in the two dimensional case, the
activation energy for the creation of the germ does not depend on the
size of the island, it is always constant : $4E$, and it stabilizes
when a row on a small facet has been removed. In the three dimensional
case, this activation energy depend on the size of the
crystallite. The transfer of a particle from a tip of the crystallite
to the germ has a gain in volume energy (depending on the size of the
islands) and a loss in edge energy of the germ (depending on the size
of the germ). Summing these two terms, an energy barrier proportional
to the size of the crystallite appears for the creation of a stable
germ. The exponential behavior of the relaxation time as a function of
N is a consequence of this energy barrier dependence.\\
Finally, we believe that the overall picture presented here,
while still oversimplified, seems to be complete enough to provide a
general picture of the processes leading to the shape relaxation of
two dimensional islands at low temperatures. \\
We acknowledge useful discussions with P. Jensen, J. Wittmer and F.
Nicaise.  We are grateful with an anonymous referee for his many
useful comments. H.L. also acknowledges partial financial support from
CONACYT and DGAPA-UNAM, and is thankful with Univ. Claude Bernard, Lyon 1,
for the invitation during which part of this work was done.

\appendix  

\section{Calculation of the probability to have 2 particles on the facet} 
\label{appen_1} 
 
We calculate the probability $P$ of having 2 particles on a facet with
absorbing boundaries, knowing that at time $t=0$, one particle is on
one edge of the facet (abscissa $1$), and that the other particle can
appear on the facet with a probability per unit time
$1/\tau$. Relatively to our Markov chain, this probability is the
probability the system in state $1$ eventually reaches state $2$.

We denote by $S(x,t)$ the probability that a particle on the facet at
position x at time $t=0$ is still on the facet at time $t$. Then,
$S(x,t)$ satisfies the usual diffusion equation:
\begin{equation}
\frac{\partial S(x,t) } {\partial t} = D \frac{\partial^2 S(x,t) } 
{\partial x^2}
\label{equa_diffu}
\end{equation}
This equation is to be solved with the conditions: $S(x,0) = 1$ for
every $x\in]0,L[$ (i.e. we are sure to find the particle on the facet
at time $t=0$), $S(0,t) = S(L,t) = 0$ for every t (i.e. the boundaries
sides of the facet are absorbent).
The solution of Equation~\ref{equa_diffu} is
\begin{equation} 
S(x,t) = \sum_{n=0}^{+\infty} \frac{4}{(2n+1) \pi} \sin(
\frac{(2n+1)\pi x}{L} ) e^{- \frac{D \pi^2 (2n+1)^2t }{L^2}}.
\end{equation} 

To take into account the appearance of particles on the facet, we assume the
process to be Poissonian so that the probability to have a particle
appearing at time $t$ is $\frac{1}{\tau} e^{-\frac{t}{\tau}}$.

Thus the probability $P$ that two particles are on
the facet is:
\begin{equation} 
P = \int_{0}^{\infty} \frac{1}{\tau} e^{-\frac{t}{\tau}} S(x,t) dt
\end{equation} 
To take into account that particles can appear on the facet on both
ends, we take $\tau = \tau_3/2$, which holds at low temperatures.
This leads to the expression :
\begin{equation} 
P = \frac{4\delta^2}{\pi} \sum_{n=0}^{+\infty} \frac{1}{2n+1} \: 
\frac{\sin((2n+1)\chi)}{\delta^2 + (2n+1)^2}
\label{appen_Po} 
\end{equation} 
where: 
\begin{eqnarray} 
\delta^2 = \frac{2L^2}{D \pi^2 \tau_3} \label{delta}\\
\chi = \frac{\pi x}{L} \label{chi}
\end{eqnarray} 

Using formula \cite{gradstein} : 
\begin{eqnarray} 
\zeta(\delta,\chi) &=& \sum_{k=1}^{+\infty} \frac{\cos(k\chi)}{\delta^2 + k^2} 
\nonumber \\
& = & \frac{\pi}{2 \delta} \frac{ \cosh (\delta(\pi-\chi))}{\sinh
(\delta \pi)} - \frac{1}{2 \delta^2}
\end{eqnarray}
we have : 
\begin{eqnarray} 
\sum_{n=0}^{+\infty} \frac{1}{2n+1} \frac{\sin( (2n+1)\chi )}
{\delta^2 + (2n+1)^2} &=& \nonumber \\ \int_{0}^{\chi} \left[
\zeta(\delta,\chi') - \frac{1}{4} \zeta(\delta/4,2\chi') \right]
d\chi'
\end{eqnarray} 
So that, using Eqs.~\ref{delta},~\ref{chi}, we finally find the following
expression for $P$:
\begin{equation} 
P = 1 - \left[\frac {2 \sinh(2 \sqrt{\rho} (L-1)) }{\sinh(2\sqrt{\rho}L) } -
\frac {\sinh(2\sqrt{\rho} (L/2-1)) }{\sinh(2\sqrt{\rho}L/2) } \right]
\label{P_exact}
\end{equation}
where $\rho=\frac{\tau_2}{\tau_3}$, and we have taken $x=1$ (the initial
particle on the face is at position 1 at time $t=0$).  We can expand
expression Eq.~\ref{P_exact} for small $\rho$ keeping the first term :
\begin{eqnarray} 
P = 2 \rho (L-1) + o(\rho)
\end{eqnarray}

\section{Probability that two particles stick on the face} 
\label{appen_11}

In this part we evaluate the probability $P_{\triangle}$ that two
particles stick on a facet with absorbing boundaries, knowing that at
time $t=0$, one particle is on one end of the facet, and the other is
at a position $x_0$ on the face. This problem can be mapped to a 2-d
problem in which, at time t, the first particle is at position y, and
the second particle is at position x. This virtual particle moves
diffusively in a square of side $L$; starting from position $(x_0,a)$
(a is the lattice spacing). The quantity we are looking for is the
probability for this virtual particle to reach the diagonal $y=x$ of
the square. Thus we can consider the motion of the virtual particle in
the triangle $ 0\leq x \leq y\leq L$.  We call $D(x,y)$ the
probability that a virtual particle starting at time $t=0$ from
$(x,y)$ leaves the triangle by the diagonal, $V(x,y)$, the probability
that this particle leaves the triangle by its vertical side, and
$H(x,y)$ the probability that this particle leaves the triangle by its
horizontal side. We use here a continuous description: the discrete
problem being far too difficult. It can be easily seen that D,V and H
satisfy Laplacian equations:
\begin{eqnarray} 
\Delta D(x,y) =0 \quad \Delta V(x,y) =0 \quad \Delta H(x,y) =0 
\end{eqnarray} 
With the conditions : 
\begin{eqnarray} 
D(x,x)=1 \quad D(x,0) = 0 \quad D(0,y) = 0 \\
V(x,x)=0 \quad V(x,0) = 0 \quad V(L,y) = 1 \\
H(x,x)=0 \quad H(x,0) = 1 \quad H(0,y) = 0 \\
 \mbox{for} \quad\forall x \in [0,L] \quad \mbox{and}\quad \forall y \in [0,L] \nonumber
\end{eqnarray}
And moreover we should have : 
\begin{equation} 
D(x,y) + V(x,y) + H(x,y) =1,
\label{probeq1}
\end{equation} 
which states that the particle is sure to leave the triangle since all
sides are absorbent.  Instead of calculating directly $D(x,y)$, we will
calculate $V(x,y)$ and $H(x,y)$. 

We first calculate $V_{\Box}(x,y)$ and $H_{\Box}(x,y)$ which are the
probability that one brownian particle in a square with absorbing
sides, respectively leaves the square by the vertical side $x=L$ and
by the horizontal side $y=0$.  So we have :
\begin{equation} 
\Delta V_{\Box}(x,y) =0 \quad \Delta H_{\Box}(x,y) =0,
\label{VHbox}
\end{equation}
with boundary conditions : 
\begin{eqnarray} 
V_{\Box}(x,L)=0  \quad V_{\Box}(x,0)=0  \\
V_{\Box}(0,y)=0  \quad V_{\Box}(L,y)=1  \\
H_{\Box}(L,y)=0  \quad H_{\Box}(0,y)=0  \\
H_{\Box}(x,0)=1  \quad H_{\Box}(x,L)=0  \\
 \mbox{for} \quad\forall x \in [0,L] \quad \mbox{and}\quad \forall y \in [0,L] \nonumber
\end{eqnarray}
The solution of equations Eq.~\ref{VHbox}, with these conditions is: 

\begin{eqnarray} 
H_{\Box}(x,y)& =& \frac{4}{\pi} \sum_{m=0}^{\infty} \frac{
\sin\frac{(2m+1)\pi x}{L}}{2m+1} \; \frac{ \sinh
\frac{(2m+1)\pi(L-y)}{L} } { \sinh (2m+1)\pi } \label{Hbox} \\
V_{\Box}(x,y)& = & \frac{4}{\pi} \sum_{m=0}^{\infty} \frac{
\sin\frac{(2m+1)\pi y}{L}}{2m+1} \; \frac{ \sinh \frac{(2m+1)\pi x}{L}
} { \sinh (2m+1)\pi } \label{Vbox}
\end{eqnarray}

One now can deduce the values of $V(x,y)$ and $H(x,y)$ of our initial
problem with a superposition of solutions to impose the condition :
$V(x,x)=0$ and $H(x,x)=0$ :
\begin{eqnarray} 
V(x,y) = V_{\Box}(x,y) - V_{\Box}(y,x) \label{vsol} \\
H(x,y) = H_{\Box}(x,y) - H_{\Box}(y,x) \label{hsol} 
\end{eqnarray} 

Using Eq.~\ref{probeq1},~\ref{vsol},~\ref{hsol},~\ref{Hbox},~\ref{Vbox},
we finally find an expression for $D(x_0,a)$ :
\begin{eqnarray}
D(x_0,a) = 1& \label{D}\nonumber \\
- \frac{4}{\pi}\sum_{m=0}^{\infty} & \frac{ \sin\frac{(2m+1)\pi a}{L}}{2m+1} \left[ \frac{\sinh \frac{(2m+1)\pi x_0}{L} - \sinh \frac{(2m+1)\pi(L-x_0)}{L}    } {\sinh (2m+1)\pi} \right]\nonumber \\
- \frac{4}{\pi}\sum_{m=0}^{\infty} &\frac{ \sin\frac{(2m+1)\pi x_0}{L}}{2m+1} \left[ \frac{\sinh \frac{(2m+1)\pi (L-a)}{L} - \sinh \frac{(2m+1)\pi a}{L}    } {\sinh (2m+1)\pi} \right] \nonumber
\end{eqnarray}

We now have to calculate the probability that the second particle is
at position $x_0$ when the other one appears on the facet. We denote
this probability $P(x_0)$. Then the probability we are looking for is
simply :
\begin{equation} 
P_{\triangle} = \int_{0}^{L} D(x_0,a) P(x_0) dx_0 \label{probDtot}
\end{equation} 
We are actually able to calculate the exact probability $P(x_0)$, but
then we are not able to find a simple expression for $P_{\triangle}$,
so we prefer to make the following approximation: in the limit of
small temperature, the typical time needed for a particle to appear on
the facet is about $\tau_3$, which is long compared to the typical
time a particle lasts on the facet (about $ L \tau_2$ for a particle
that starts near the edge). Thus, to a good approximation, the
probability $P(x_0)$ has reached its stationary value. Taking into
account only the first term of the series, we have:
\begin{equation} 
P(x_0) = \frac{\pi}{2L} \sin \frac{\pi x_0}{L}
\label{P(x0)}
\end{equation}
With this expression Eq.~\ref{probDtot} becomes easy to calculate and we find :
\begin{eqnarray} 
P_{\triangle} & = & 1 - \frac{ \sinh \frac{\pi (L-a)}{L} - \sinh
\frac{\pi a}{L} }{\sinh \pi} \label{Pprems}\\ & = & \left(\frac{\cosh
\pi +1 } { \sinh \pi} \right) \frac{\pi a}{L} - \frac{\pi^2 a^2}{2L^2}
+ o(1/L^2) \label{Pdeux}
\end{eqnarray} 
So that we find that the leading term of $P_{\triangle}$ is
proportional to $1/L$. We write this as:
\begin{eqnarray} 
P_{\triangle} = \lambda \frac{1}{L} + o(1/L^2) \label{appen11_final}\\
\mbox{with} \; \lambda = \left(\frac{\cosh \pi +1 } { \sinh \pi} \right)\pi
\end{eqnarray}

\section{Calculation of the time to leave state 2}
\label{appen_12}

We calculate in this part the average time $<\tau>$ for two particles
on the facet either to bond, or either one of them to reach the
boundary. This time corresponds to the time the system stays in
state $2$ in the Markov chain. \\
Using the description of appendix~\ref{appen_11} with the virtual
particle in the triangle, we are looking for the average time this
particle needs to leave the triangle.

Let us call $S(x_0,y_0,t)$ the survival
probability: i.e. the probability a particle starting at
$(x_0,y_0)$ at time $t=0$ is still in the triangle at time t. The
average time $<\tau (x_0,y_0) >$ the particle stays in the triangle is
given by :
\begin{equation} 
<\tau (x_0,y_0) > = \int_{0}^{\infty} S(x_0,y_0,t) dt
\label{detau}
\end{equation} 
Moreover calling $P_{\triangle}(x,y,x_0,y_0,t)$ the probability that
one particle starting at $(x_0,y_0)$ at time $t=0$ is at position
$(x,y)$ at time $t$, we have :
\begin{equation} 
S(x_0,y_0,t) = \int \!\!\!\! \int_{\triangle}
P_{\triangle}(x,y,x_0,y_0,t) dxdy
\label{defS} 
\end{equation}
Finally, we are only interested in $<\tau (x_0,1)>$ since we know that
our virtual particle starts at $(x_0,a)$, so we have to average the time
$<\tau (x_0,1)>$ over all values $x_0$. To do this we use the approximate
distribution given in Eq.~\ref{P(x0)} :
\begin{equation} 
<\tau> = \int_{0}^{L} P(x_0)  <\tau (x_0,1) > dx_0 
\label{deftau} 
\end{equation} 
So finally, using Eq.~\ref{detau},~\ref{defS},~\ref{deftau}, we need
to evaluate:
\begin{equation}
<\tau> = \int_{0}^{L} \int_{0}^{\infty} \int \!\!\!\! \int_{\triangle}
P(x_0) P_{\triangle}(x,y,x_0,1,t) dxdy\; dt \; dx_0
\label{tautotal} 
\end{equation}

We now have to calculate $ P_{\triangle}(x,y,x_0,1,t)$. As before, we
find the solution in a square, and then by superposition,
we deduce the solution in the triangle :
\begin{eqnarray} 
&P_{\triangle}(x,y,x_0,1,t) = P_{\Box}(x,y,x_0,1,t) -
P_{\Box}(x,y,1,x_0,t) \label{Ptri}\\ &P_{\Box}(x,y,x_0,1,t) =
\nonumber \\ &\frac{4}{L^2} \sum_{m,n=1}^{\infty}
\sin\frac{m \pi x_0}{L}\sin\frac{n \pi}{L} \sin\frac{m \pi x}{L} \sin\frac{n \pi y}{L} e^{-\frac{D(m^2+n^2) \pi^2 t}{L^2}} \label{Pbox}
\end{eqnarray} 
We find $<\tau>$ integrating Eq.~\ref{tautotal} using Eq.~\ref{Ptri},~\ref{Pbox} over $x_0$ first, and then over $x$ and $y$, and finally
over $t$, we find :
\begin{equation}
<\tau> = \frac{L^2}{D \pi^3} \sum_{k=1}^{\infty} \frac{1}{1+4k^2} \; \sin\frac{2k \pi}{L} \; \left[ \frac{2}{k} - \frac{1}{k(2k+1)} \right] \label{ztau}
\end{equation} 
We interested in the leading term as $L \rightarrow \infty$, let us define
a function $f(u)$ by :
\begin{equation} 
f(u) = \sum_{k=1}^{\infty} \frac{1}{1+4k^2} \; \sin(ku) \; \left[ \frac{2}{k} - \frac{1}{k(2k+1)} \right]
\end{equation}
$f(u)$ is a normally convergent series, so that : 
\begin{eqnarray} 
\frac{df(u)}{du} &=& \sum_{k=1}^{\infty} \frac{1}{1+4k^2} \; \cos(ku)
\; \left[2 - \frac{1}{(2k+1)} \right] \\
                 &\stackrel{u\rightarrow 0}{\longmapsto}&
\sum_{k=1}^{\infty} \frac{1}{1+4k^2} \; \left[2 - \frac{1}{(2k+1)}
\right] \label{premterm}
\end{eqnarray}
which is also convergent. The second term of the development is of order
$u^2$ or smaller, so that integrating Eq.~\ref{premterm},
and using it in Eq.~\ref{ztau}, we find:
\begin{equation}
<\tau> = \frac{2L}{D \pi^2} \sum_{k=1}^{\infty} \frac{1}{1+4k^2} \; \;
\left[ 2 - \frac{1}{(2k+1)} \right] + O(1/L)
\end{equation} 
 
Thus $<\tau>$ is proportional to L, and we write: 
\begin{eqnarray} 
<\tau> = \kappa L \tau_2 + O(1/L) \label{appen12_final}\\
\mbox{ with} \; \kappa = \frac{4}{\pi^2} \sum_{k=1}^{\infty} \frac{1}{1+4k^2} \;  \; \left[ 2 - \frac{1}{(2k+1)} \right]
\end{eqnarray}

\section{Calculation of the time a particle needs to go to center of the facet}
\label{appen_2} 

We now calculate the average real time $t_{p,q,\alpha}$ to go from
state $i$ to state $i+1$ assuming the average distance between the
kinks or corners from which particles are emitted and the germ is
$L/2$. This again can be posed as a Markov chain where :\\
- state $0$ : no particle is on the facet;\\
- state $1$ : one particle coming from a kink is in position 1 on the facet;\\
- state $2$ : the particle is at position $2$; \\
\ldots \\
- state $k$ : the particle is at position k; 
and state $L/2$ is absorbent. 

Again we use $\tau_2$ as the unit time. Using the
same definition for the quantities $p_i$, $q_i$ and $\alpha_i$ as in
the main text, we have:
\begin{eqnarray} 
p_i &=& 1/2 \\
\alpha_i&=&0 \\
q_i &=& 1/2 \\
 with & & i \ge 1 \nonumber
\end{eqnarray} 
To calculate $p_0$ and $\alpha_0$ we know that the real time an atom
needs to leave a kink is $\tau_3$, and as there are two kinks at the
ends of the facet, $\tau_3/2$ is, to a good approximation, the average
time to leave state $0$. Thus that we find:
\begin{eqnarray} 
p_0 &=& 2\rho \\
\alpha_0 &=& 1 - 2\rho 
\end{eqnarray}
Calling $n_i$ the average time to go from state $i$ to the absorbent
state $L/2$, we have the equations :
\begin{eqnarray} 
n_0 &=& 1 + \alpha_0 n_0 + p_0 n_1 \label{appen_eq0} \\ 
n_k &=& 1 + \frac{n_{k-1}}{2} + \frac{ n_{k+1}}{2} \label{appen_eqk} \\
 with & & k \ge 1 \nonumber\\ 
n_{L/2}&=& 0 \label{appen_eqL}
\end{eqnarray} 
and $n_{L/2}=0$. \ Summing equations~\ref{appen_eqk} from $k=1$ to
$j$, and then from $j=1$ to $L/2-1$, and using Eq.~\ref{appen_eqL}, one
finds :
\begin{equation} 
n_1 = \frac{L-2}{L} n_0 + \frac{L-2}{2} 
\end{equation} 
And Eq.~\ref{appen_eq0} permits to obtain $n_0$ :
\begin{equation} 
n_0 = \frac{L}{2p_0} + \frac{L(L-2)}{4} 
\end{equation} 
Going back to real time, one finds the average time $\tau$ a particle
needs to leave a kink and reach the center of the facet is:
\begin{equation} 
t_{p,q,\alpha} = \frac{L}{4}\tau_3 + \frac{L(L-2)}{4} \tau_2 
\label{appen_2_final}
\end{equation}

\section{Calculation of $n_0$}
\label{appen_3}

To carry out the calculation, we can first calculate the $n_k$ for $k
\ge 1$.  Noting that $p=q$, we have from Eq.~\ref{eqk}:
\begin{equation}
(n_{k+1}-n_k) = (n_k - n_{k-1}) - 1/p,
\end{equation} 
so that summing  from k=4 to j, and then from j=4 to
$l$, and using $n_{l+1}=0$, we find:
\begin{equation}
(l-1) n_3 = (l-2) n_2 + 1/p( \frac{(l-2)(l-1)}{2})  
\end{equation}
Inserting this result in Eq.~\ref{eq2} we find $n_2$ as a function
of $n_1$, in the same way, using Eq.~\ref{eq1}, one can deduce the
equation for $n_1$ :
\begin{equation} 
n_1 \left( 1- \alpha_1 - p_1 B \right) = 1+p_1 A + q_1 n_0 \label{n1}
\end{equation}
where : 
\begin{eqnarray}
A &= &\frac{ 1 +\frac{p_2}{p} \frac {l-2}{2} } { 1 - \alpha_2 - p_2
\frac{l-2}{l-1} } \label{A}\\ B &=& \frac { q_2} { 1 - \alpha_2 - p_2
\frac{l-2}{l-1} } \label{B}\\
\end{eqnarray}

We now calculate the case $k < 0$.\\ Calling $m_j=n_{-j}$ for every
$j$, and summing Eq.~\ref{eq-k} from j to $z-1$, one finds :
\begin{equation}
m_z-m_{z-1}=m_{j}-m_{j-1}-\frac{z-j}{p^*} \label{mz},
\end{equation} 
and Eq.~\ref{eq-l} yields: $m_l-m_{l-1}=\frac{1}{2p^*}$. Using this in
Eq.~\ref{mz} and taking $j=3$, one finds :
\begin{equation} 
m_3-m_2=\frac{l-3}{p^*} + \frac{1}{2p^*} 
\end{equation} 
Using Eqs.~\ref{eq-2},~\ref{eq-1}, one finds :
\begin{equation}
n_{-1}= \frac{1}{q_1^*} + \frac{p_1^*}{q_1^* q_2^*} + \frac
{p_1^*p_2^*}{q_1^* q_2^* p^*} (l-5/2) + n_0 \label{n-1}.
\end{equation} 

One can know obtain the value of $n_0$ from Eq.~\ref{n1},~\ref{n-1} and~\ref{eq0} :
\begin{eqnarray} 
\lefteqn{ p_0(1-\frac{q_1}{1- \alpha_1 - p_1 B}) n_0 =} \nonumber \\
& & 1+ p_0 \frac{1+p_1 A}{1- \alpha_1 - p_1 B} \nonumber \\ & & + q_0
\left( \frac{1}{q_1^*} + \frac{p_1^*}{q_1^* q_2^*} + \frac
{p_1^*p_2^*}{q_1^* q_2^* p^*} (l-5/2) \right)
\label{n0_exact}
\end{eqnarray}

Then, using
Eq. [\ref{p1}-\ref{q0}] we can calculate A and B :
\begin{eqnarray*}
A & = & \frac{L}{4 \rho} \frac{(l-1)(l-2)}{(\frac{L}{\lambda}-1)(l-1)+1} \nonumber \\
& & + \frac{ \frac{\kappa}{\lambda} L^2(l-1)+
\frac{L(L-2)(l-2)(l-1)}{4} }{(\frac{L}{\lambda}-1)(l-1)+1} + o (1)\\ 
B & = & 1 -
\frac{1}{(\frac{L}{\lambda}-1)(l-1)+1}
\end{eqnarray*}
And finally, the expression of $n_0$ is: 
\begin{eqnarray}
n_0 &=& \frac{1}{\rho^2}\frac{(L+2l)}{4L(L-1)} \left[
 (\frac{L}{\lambda}-1)(l-1)+1 \right] \nonumber \\ & & + O(1/\rho)
\label{no_total}
\end{eqnarray}

We report here only the leading term: the expressions of the
quantities $p_i,q_i$ and $\alpha_i$ permit us to calculate $n_0$ up
to order $1$, but these terms are truly ugly and it does not seem
relevant to give them here.

Going back to real time, we find that the time $\tau(L,l)$ to
nucleate a new row on a facet is given by :
\begin{eqnarray}
\tau(L,l) &=& \frac{\tau_3^2}{\tau_2} \frac{(L+2l)}{4L(L-1)} \left[ (\frac{L}{\lambda}-1)(l-1)+1 \right] \nonumber \\
 & & +  O(\tau_3)
\label{tau_total}
\end{eqnarray}


\begin{thebibliography}{99}
\bibitem{note} 
This formulation of the relaxation time as a function
  of $N/N_c$ permits to resolve the apparent paradox mentioned in
  \cite{eur_physB}. Namely, that extrapolating the low temperatures
  regime to higher temperatures, one could infer that large
  islands would relax faster at high temperature than at low
  temperature. The curve $log(t_{relaxation})$ as a function of
  $log(N)$ at a given temperature $T_1$ is obtained by translation of
  the curve at temperature $T_0$ by a vector
  $\frac{N_c(T_1)}{N_c(T_0)} \left(\begin{array}{c}1\\5 \end{array}
  \right)$. Since the slope of this translation is greater than the
  slopes of $log(t_{relaxation})$ as a function of $log(N)$, thus two
  curves at different temperatures can not cross each other.
\end{thebibliography}
\end{document}